\documentclass[aps,pra,twocolumn,superscriptaddress]{revtex4-2}
\usepackage{amsmath,amsfonts,amssymb}
\usepackage{graphicx,float,calc}
\usepackage{color,bm}
\usepackage{ulem}
\usepackage{braket}
\usepackage[colorlinks,urlcolor=blue,citecolor=blue,linkcolor=blue]{hyperref}

\begin{document}
\title{Discrete time crystals enhanced by Stark potentials in Rydberg atom arrays}
	
\author{Jian-Jia Wang}
\affiliation{Key Laboratory of Atomic and Subatomic Structure and Quantum Control (Ministry of Education), School of Physics, South China Normal University, Guangzhou 510006, China}
\affiliation{Guangdong-Hong Kong Joint Laboratory of Quantum Matter, Frontier Research Institute for Physics, South China Normal University, Guangzhou 510006, China}

\author{Ling-Zhi Tang}
\thanks{tanglingzhi@quantumsc.cn}
\affiliation{Quantum Science Center of Guangdong-Hong Kong-Macao Greater Bay Area (Guangdong), Shenzhen 518045, China}

\author{Yan-Xiong Du}
\affiliation{Key Laboratory of Atomic and Subatomic Structure and Quantum Control (Ministry of Education), School of Physics, South China Normal University, Guangzhou 510006, China}	
\affiliation{Quantum Science Center of Guangdong-Hong Kong-Macao Greater Bay Area (Guangdong), Shenzhen 518045, China}

\author{Dan-Wei Zhang}\thanks{danweizhang@m.scnu.edu.cn}
\affiliation{Key Laboratory of Atomic and Subatomic Structure and Quantum Control (Ministry of Education), School of Physics, South China Normal University, Guangzhou 510006, China}
\affiliation{Guangdong-Hong Kong Joint Laboratory of Quantum Matter, Frontier Research Institute for Physics, South China Normal University, Guangzhou 510006, China}
\affiliation{Quantum Science Center of Guangdong-Hong Kong-Macao Greater Bay Area (Guangdong), Shenzhen 518045, China}

\begin{abstract}
Discrete time crystals (DTCs) are non-equilibrium phases in periodically driven systems that exhibit spontaneous breaking of discrete time-translation symmetry. The stabilization of most DTC phases is achieved via the disorder-induced many-body localization. In this work, we propose an experimental scheme to realize disorder-free DTCs in a periodically driven Rydberg atom array. Our scheme utilizes a linear potential in the atomic detuning to enhance the DTC order, {without being tired to (Stark) many-body localization.} We numerically demonstrate that the Stark potential enhances the robustness of the DTC against the flip imperfections and extends its lifetime, which are independent of initial states. Thus, our scheme provides a promising way to explore DTCs in Rydberg atom arrays without disorder averaging and special state preparation.
\end{abstract}
	
\date{\today}
\maketitle
\section{INTRODUCTION}
The concept of time crystals that spontaneously break the continuous time translation symmetry was originally proposed by Wilczek \cite{Wilczek2012}. Although the realization of continuum time crystals in quantum ground states or equilibrium states has been ruled out \cite{Bruno2013,Watanabe2015}, the discrete time crystals (DTCs) with spontaneous breaking of discrete time-translation symmetry in non-equilibrium periodically driven (Floquet) systems were proposed \cite{Sacha2015,Khemani2016,Else2016,Yao2017,Zaletel2023Review,Khemani2019}. The properties of DTCs have been extensively studied \cite{Yao2017,Russomanno2017,Huang2018,Yu2019,Sacha2017,Gambetta2019,Fan2020,Giedrius2021,Collura2022,Kshetrimayum2020,Kshetrimayum2021,Liu2023,Chandra2024}. In a generic Floquet system without local conservations, periodic driving makes the system evolving toward an infinite-temperature state via unbounded heating \cite{Pedro2015}, and thus the DTC phase cannot be sustained in the long time limit. Several mechanisms have been suggested to prevent thermalization to enable long-lived DTCs \cite{Yao2017,Kshetrimayum2020,Kshetrimayum2021,Liu2023,Choi2017,Rovny2018PRL,Zhang2017,Mi2022,Frey2022,Tang2021,Tang2025,Li2024,Zhang2020,Liang2023,Ni2021}, such as many-body localization (MBL) with strong disorders \cite{Abanin2019,Sierant2025}, quantum many-body scars for special initial states \cite{Maskara2021,Huang2022} and prethermal mechanism \cite{Else2017,Pizzi2021,kyprianidis2021,Luitz2020,Stasiuk2023,Beatrez2023,CYing2022}.
Although there is no dependence of initial states in the MBL-DTC systems, one requires many disorder configurations for averaging over observables to characterize DTC orders. Recently, it was found that a strong Stark potential in many-body systems can suppress thermalization and lead to localization of generic initial states without disorders, termed as Stark MBL \cite{vanNieuwenburg2019,Schulz2019,Kshetrimayum2020,Doggen2021}. {The Stark MBL with dynamical constricts of excitations can suppress heating to an infinite-temperature state, which provides a new way to stabilize DTCs in clean Floquet many-body systems \cite{Liu2023,BarLev2024}. Notably, the linear Zeeman field can be effectively suppressed by the spin-flip pulses over two sequent periods \cite{Khemani2019}, and thereby a linear (unbounded) Ising interaction term is needed to realize the Stark MBL-DTC phase \cite{Liu2023,BarLev2024}.}

{In recent years, DTC orders have been observed experimentally in various artificial quantum systems \cite{Choi2017,Zhang2017,Rovny2018PRL,Rovny2018PRB,Autti2018,Pal2018,Smits2018,Giergiel2020,Smits2021,Mi2022,Frey2022,Stasiuk2023,Beatrez2023,CYing2022,Liu2024,Wang2021,DeRoeck2017,Ho2017}, such as dipolar nitrogen vacancy ensembles \cite{Choi2017} and nuclear magnetic resonances \cite{Rovny2018PRL} without MBL, trapped ions \cite{Zhang2017} and superconducting qubits \cite{Mi2022,Frey2022} in the MBL regime. Despite the experimental progress, realizing the MBL-DTC phase remains an outstanding challenge, as stringent conditions (strong disorders and short-range interactions) are required to achieve Floquet MBL \cite{Zaletel2023Review,Khemani2019,Abanin2019,Sierant2025} and the MBL phase could be unstable against the avalanche effect \cite{Sierant2025,Leonard2023}. Based on prethermalization mechanism \cite{Luitz2020}, long-lived DTC responses without disorders have also been observed in these systems \cite{Stasiuk2023,Beatrez2023,CYing2022}.} On the other hand, Rydberg atom arrays have become an important experimental platform to explore non-equilibrium many-body physics \cite{Labuhn2016,Barredo2016,Bernien2017,Lin2019PRL,Lin2020PRB,Ebadi2021,Scholl2021,Zhang2018}.  Rydberg atoms can be individually addressed in the array with locally tunable parameters, through holographic optical techniques enabling site-specific manipulation of excitation fields and interaction geometries. The Rydberg state has a long lifetime to enables the study of coherent dynamics. By selecting different Rydberg levels, diverse interactions and potentials can be engineered to simulate various models \cite{Kim2016,Barredo2015,Nguyen2018,Bernien2017,Fan2020,Wu2021review}. Remarkably, the DTC order of quantum many-body scars has been observed in a one-dimensional array of periodically driven Rydberg atoms \cite{Bluvstein2021}. {An interesting problem in this atomic system is how to achieve the DTC order without the requirement of MBL and scar states.}

In this work, we propose an experimentally feasible scheme with a Floquet model to realize clean DTCs in a driven Rydberg atom array. Our scheme utilizes a Stark potential in the atomic detuning to stabilize the DTC order. {This term, acting as a linear Zeeman field, does not lead to Stark MBL under periodical spin flips \cite{Liu2023,BarLev2024}, but can effectively induce prethermal regimes with approximate U(1) symmetry for long-lived DTC orders \cite{Luitz2020,Stasiuk2023,Beatrez2023,CYing2022}.} By numerically computing various dynamical observables in a finite Rydberg atom array with open boundaries, we demonstrate that the Stark potential enhances the robustness of the DTC against the flip imperfections. We also find that the Stark potential can extend the lifetime of the DTC under imperfections. The presence and enhancement of the DTC order and lifetime via the Stark potential are independent on initial states. Thus, our present scheme provides a promising way to explore DTCs with much fewer resources in Rydberg atom arrays, as it does not require disorder averaging and special state preparation.

The rest of paper is organized as follows. Sec.~\ref{sec:model} introduces the model and scheme with a Rydberg atom array to realize the clean DTCs. In Sec.~\ref{sec:results and dicussion}, we numerically show that the Stark potential enhances the robustness of the DTC order against the flip imperfections and prolongs its lifetime for different initial states. Finally, a short conclusion is presented in Sec.~\ref{sec:conclusion and outlook}.

\section{MODEL AND SCHEME}
\label{sec:model}

\begin{figure}[tb]
	\centering
	\includegraphics[width=0.46\textwidth]{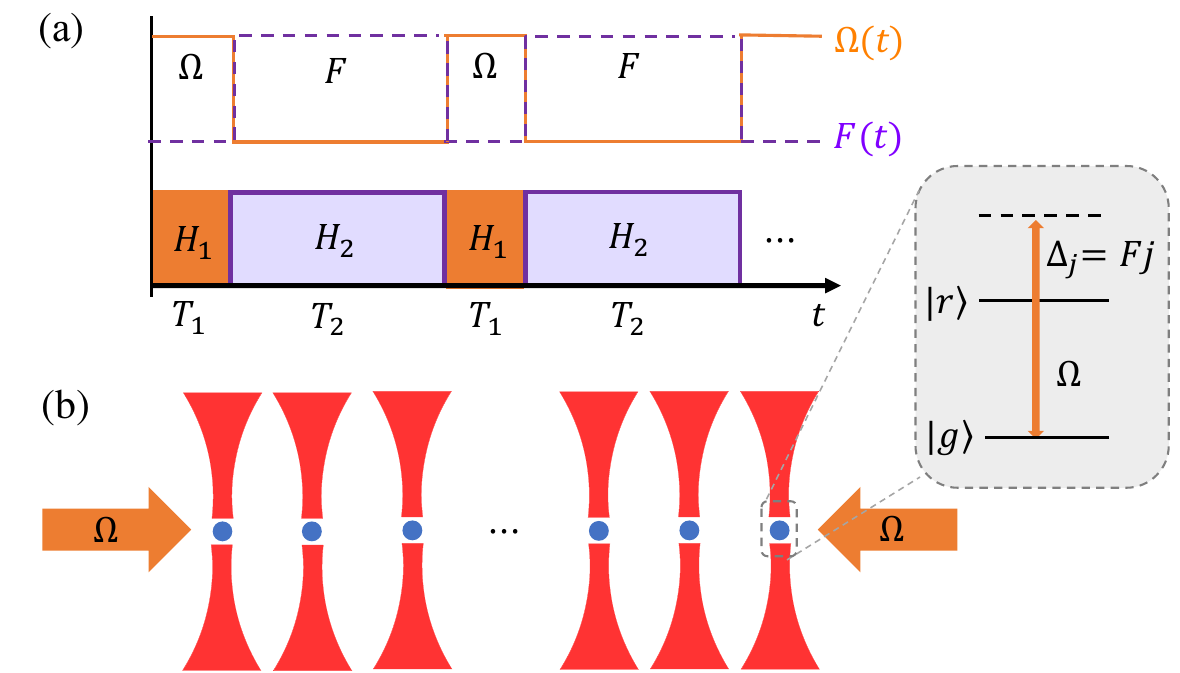}
	\caption{Experimental scheme. (a) Sequence diagram for periodically switching on the Raman field $\Omega$ and Stark potential $\Omega$ in stages $T_1$ and $T_2$, respectively. The sequence corresponds to the Hamiltonians $H_1$ and $H_2$ in the two stages. (b) Schematic of a one-dimensional $^{87}$Rb atom arrays. The horizontal orange laser is the Raman laser, which acts at the $T_1$ stage with a Rabi frequency $\Omega$. The vertical red laser is the optical tweezers, which are used to trap atoms and act as a Stark potential field from the $T_2$ stage. Atom at the j-th site is initialized in the ground state $\ket{r}$ and coupled to the Rydberg state $\ket{r}$ via a two-photon Raman transition with a site-dependent and linear detuning $\Delta_j=Fj$.}\label{fig1}
\end{figure}

We consider a one-dimensional array of Rydberg atoms trapped by optical tweezers with open boundaries \cite{Kim2016,Bernien2017}, as illustrated in Fig. \ref{fig1}. The optical tweezer array is created by a spatial light modulators (SLM). Each Rydberg atom can be viewed as a two-level (spin-1/2) system comprising an electronic ground state $\ket{g}$ and a high-lying Rydberg state $\ket{r}$, with a transition frequency $\omega_{gr}$ between the two states. In experiments, the two atomic states are coupled via a two-photon Raman process with site- and time-dependent Rabi frequency and detuning. We consider homogeneous Rabi frequency $\Omega(t)$ and spatially various detuning $\Delta_j(t)=\Delta(t)+\delta_j(t)$. Here $\Delta(t)=\omega_d-\omega_{gr}$ is a global term with $\omega_d$ as the frequency of the Raman beams, and $\delta_j(t)$ denotes the $j$-site (atom) detuning, which has been experimentally realized by site-dependent Stark shifts of a second SLM \cite{Manovitz2025}. The interaction between both two atoms at $i$ and $j$ sites exists when they are in the Rydberg states and takes the van der Waals form $V(R)=C_6/R^6$, where $C_6$ is the vdW coefficient and $R=|i-j|$ denotes the interatomic distance.
We focus on the nearest-neighbor interaction and will show that longer-range interaction terms are negligible. In the rotating frame and let $\hbar=1$, the system Hamiltonian can be written as \cite{Manovitz2025}
\begin{equation}
  \begin{split} \label{Ht}
 H(t) = \Omega(t)\sum_{j=1}^{L} \sigma^x_j+\sum_{j=1}^{L}\Delta_j(t) n_j+\sum_{j=1}^{L-1}V n_j n_{j+1},
  \end{split}
\end{equation}
where $\sigma^x_j=\ket{r}_{jj}\bra{g}+\text{h.c.}$, $n_j=\ket{r}_{jj}\bra{r}$, and $L$ is the system size.

We now propose a Floquet protocol to realize DTC orders in this system, as illustrated in Fig. \ref{fig1}(a). The protocol comprises two parts: $H_1$ and $H_2$ respectively correspond to the first ($T_1$) and second ($T_2$) stages of a Floquet cycle with period $T=T_1+T_2$. During the $T_1$ stage with $t\in[0,T_1]$, the light field is resonantly turned on with a constant Rabi frequency $\Omega(t)=\Omega$ and vanishing global detuning $\Delta(t)=0$ without the second SLM, such that $\Delta_j(t)=0$ in this stage. During the $T_2$ stage with $t\in [T_1, T_1+T_2]$, the light field is switched off with $\Omega(t)=0$ and only the second SLM is turn on with $\Delta_j(t)=\delta_j(t)=Fj$, which is a linear Stark potential with the strength $F$ as shown in Fig. \ref{fig1}(b). The corresponding Floquet unitary operator for a period $T=T_1+T_2$ is given by 
\begin{equation}
		U_F=U_2U_1=e^{-iH_2T_2}e^{-iH_1T_1},
\end{equation}
where
\begin{align}
		& H_1 = \sum_{j=1}^{L} (\Omega+\epsilon)\sigma^x_j+\sum_{j=1}^{L-1}V n_j n_{j+1}, \label{H1}\\
	    & H_2 = \sum_{j=1}^{L-1}V n_j n_{j+1} + \sum_{j=1}^{L}Fj n_j.                                      \label{H2}
\end{align}

Here we add an imperfection $\epsilon$ into the spin flip term with fixed $\Omega T_1=\pi/2$, which quantifies the deviation from the ideal Rabi frequency and is used to show the rigidity of the DTC order \cite{Yao2017}. Without the imperfection and interaction, $U_1$ term flips spins around the X-axis to the opposite polarization when $\Omega T_1=(2n+1)\pi/{2}$, which corresponds to a $2T$-period response of the DTC order. {Notably, different from the Stark MBL-DTC case where a strong Stark term in the interaction part is added \cite{Liu2023,BarLev2024}, we here put the Stark potential to the single-spin term in Eq. (\ref{H2}). This term acting as an effective linear Zeeman field is suppressed under spin-flip pulses (echoed out under perfect $\pi$-flip to leading order \cite{Luitz2020}), and thereby cannot induce MBL type of effects. Instead, such a strong linear field in Floquet settings can realize an approximate long-lived U(1) conservation under imperfection flips \cite{Luitz2020}, which is approximately the total spin along the $z$ direction (or say, the global Rydberg excitation in the Rydberg atom array). Thus, our model does not realize Stark 
MBL-DTCs but rather is relevant to the U(1) prethermal DTCs \cite{Luitz2020,Stasiuk2023,Beatrez2023,CYing2022}. The long-lived DTC responses enabled by the Stark field for Floquet Rydberg atoms are numerically shown in the following section.} We consider moderate interaction strength $VT_1$ and keep the same magnitudes of $\Omega T_1$ and $VT_2$ to demonstrate the DTC response enabled by the Stark potential. In our numerical simulations, we choose $\epsilon T_1$, $VT_1$, $VT_2$ and $FT_2$ as dimensionless parameters by setting $T_1=T_2/10=1$.

The scheme can be implemented with the existing technology \cite{Barredo2016,Bernien2017}. One can first probabilistically load $^{87}$Rb atoms from a cold atomic cloud into a one-dimensional optical tweezer array created by a SLM. Defects in the array are identified using a real-time feedback imaging system and corrected by transporting additional $^{87}$Rb atoms to empty sites with a pair of orthogonally aligned acousto-optic deflectors. In typical experiments, the configuration of each atom is a three-level ladder structure, where a two-photon process nearly resonantly couples the ground state $\ket{g}=\ket{5S_{1/2},F=2,m_F=2}$ to the Rydberg state $\ket{r}=\ket{50S_{1/2},j=1/2,m_j=1/2}$ via intermediate state $\ket{e}=\ket{5P_{1/2},F=2,m_F=1}$. The Stark potential then can be site-selectively engineered by incorporating second SLM during the $T_2$ stage, which generates position-dependent light shifts to controllably modify atomic energy levels at individual lattice sites \cite{Manovitz2025}. {The challenge to observe the DTC response is the limited many-body coherence time in practical experiments. A strategy to have more driving periods within the coherence time is increasing the Rabi frequency $\Omega$. By choosing $\Omega/2\pi=25$ MHz and $T_2=10T_1=0.1~\mu$s, for instance, one can have dozens of $2T$-periods ($T=0.11~\mu$s) of DTC responses for typical coherence timescale of several microseconds \cite{Bluvstein2021,Manovitz2025}. Different from the many-body scar case in the Rydberg blockade regime \cite{Bluvstein2021} with interesting strength $V\gg\Omega$ in Eq. (\ref{H1}), we instead take  $V\approx0.1\Omega$ in our numerical simulations, which is sufficiently large for the DTC order under the Stark potential.}

\begin{figure}[htb]
    \centering
	\includegraphics[width=0.46\textwidth]{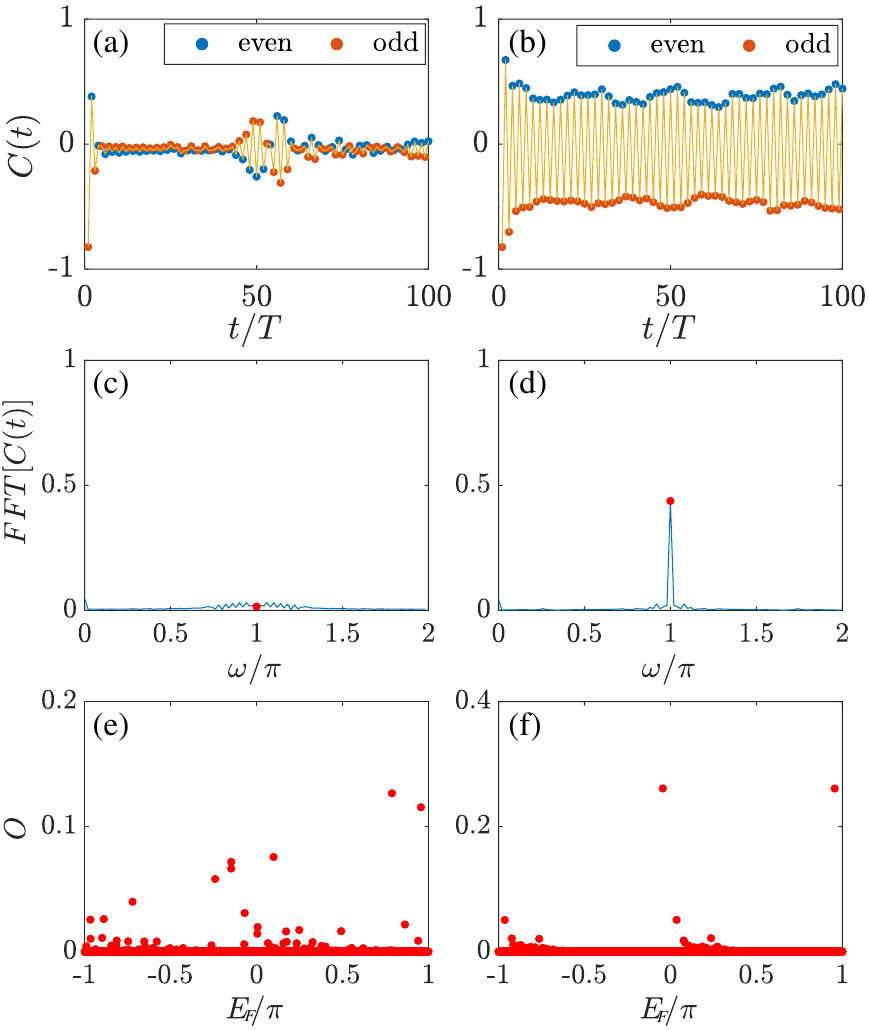}
	\caption{The autocorrelator $C$ as a function of time $t/T$ for (a) $FT_2=0$ (absence of Stark potentials) and (b) $FT_2=0.25$. (c) and (d) Fourier spectra $FFT[C(t)]$ corresponding to (a) and (b), respectively. (e) and (f) Overlap $O$ between the initial state and quasi-eigenstates with respect to the quasi-eigenenergy $E_F$, which correspond to (a) and (b), respectively. The initial state in (a-f) is $|\psi(0)\rangle=\ket{111111111111}$. Other parameters are $T_1=1$, $T_2=10$, $V=0.1$, $\epsilon=0.3$, and $L=12$.}\label{fig2}
\end{figure}

\section{Numerical results}
\label{sec:results and dicussion}

In this section, we numerically study the clean DTC enhanced by the Stark potential in the atom array of size $L$ via the exact diagonalization. To characterize the DTC response, we employ the site-averaged spin autocorrelator $C(t)$ and its fast Fourier transform $FFT[C(t)]$, which are dynamical observables in atom arrays \cite{Yao2017,Liu2023,Mi2022}. The autocorrelator is given by
\begin{equation}	C(t)=\overline{\left\langle{Z(0)Z(nT)}\right\rangle}=\frac{1}{L}\sum_{j=1}^{L}\left\langle{Z_j(0)Z_j(nT)}\right\rangle,
\end{equation}
where $t=nT$, $Z_j(nT)=(U^n)^{\dag}\sigma_j^{z}U^n$ with $\sigma_j^{z}=\ket{r}_{jj}\bra{r}-\ket{g}_{jj}\bra{g}$. In sufficiently large Floquet cycles, the characteristics of DTC can be assessed by observing the Fourier spectrum of the autocorrelators \cite{Yao2017}:
\begin{align}
       FFT[C(t)]=\sum_{n=1}^{N}C(t)e^{-i\frac{2\pi}{N} \omega n},
\end{align}
where $N$ is the number of Floquet cycles, and $\omega$ denotes the frequency in unit of $1/T$. {Here we choose $N=100$ in our numerical simulations, which is large enough to obtain the Fourier spectrum. }

\begin{figure}[htp]
	\centering
	\includegraphics[width=0.48\textwidth]{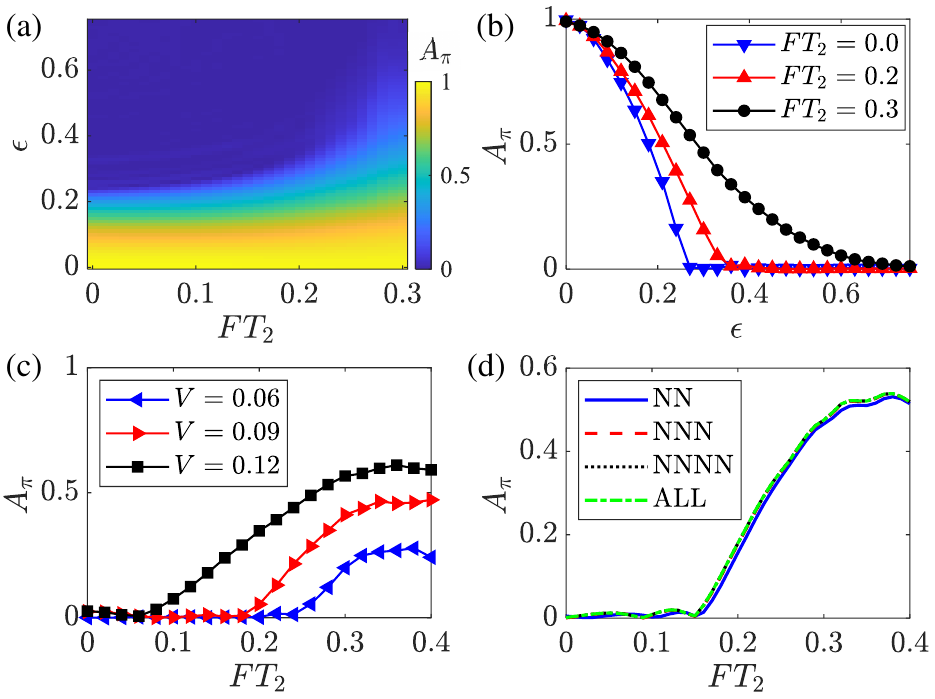}
	\caption{(a) Amplitude of the subharmonic response $A_{\pi}$ as functions of $\epsilon$ and $FT_2$ for $V = 0.1$. (b) $A_{\pi}$ as a function of $\epsilon$ for $FT_2=0,0.2,0.3$ and fixed $V = 0.1$. (c) $A_{\pi}$ as a function of $FT_2$ for $V=0.06,0.09,0.12$ and fixed $\epsilon=0.3$. (d) $A_{\pi}$ as a function of $FT_2$ with nearest-neighbor interaction (NN), next-nearest-neighbor (NNN), next-next-nearest-neighbor (NNNN) and every atom interacting with others (ALL). Other parameters in (a-d) are $T_1=1$, $T_2=10$ and $L=10$, and the initial product states is $\ket{111111111111}$.}\label{fig3}
\end{figure}

We first show that the Stark potential can enhance the DTC under imperfection $\epsilon$. Typical numerical results of for large imperfection with $\epsilon=0.3$ are shown in Fig. \ref{fig2}. In the absence of the Stark potential with $FT_2=0$, there is no DTC response in Fig. \ref{fig2}(a). Under the Stark potential with $FT_2=0.25$, one can see the $2T$-period of $C(t)$ as the DTC response in Fig. \ref{fig2}(b). To be more clearly, we plot the corresponding Fourier spectra $FFT[C(t)]$ in Figs. \ref{fig2}(c) and \ref{fig2}(d), where a peak is absent and present at $\omega=\pi$. We can use the amplitude of the peak as $A_{\pi}=FFT[C(t)]|_{\omega=\pi}$ to further reveal the DTC order with subharmonic periodic oscillatory and finite value of $A_{\pi}$. We also numerically compute the overlap between the initial state $\ket{\psi(0)}$ and the quasi-eigenstates of the Floquet operator $U_F$, which is given by
\begin{equation}
    	O=|\langle{\psi(0)}\ket{\psi^F_\alpha}|^2.
\end{equation}
Here the Floquet eigenstates $\ket{\psi^F_\alpha}$ with eigenenergies $E_{F}$ are obtained by solving the equation $U_F\ket{\psi^F_\alpha}=e^{-iE_{F}}\ket{\psi^F_\alpha}$. The results of the overlap $O$ with respect to the quasi-energies $E_{F}$ are shown in Figs. \ref{fig2}(e) and \ref{fig2}(f) for $FT_2=0$ and $FT_2=0.25$, respectively. One can see a pair of Floquet eigenstates with quasi-energy difference of $\pi$ are exhibited under the Stark potential in Fig. \ref{fig2}(f), which is termed as a $\pi$ pair and corresponds to the stable subharmonic response \cite{Liu2023}. {The overlap for the dominant $\pi$-pair eigenstates and other eigenstates is increased and decreased by increasing the strength of the Stark potential (not shown here), respectively. This implies that the Stark field suppress the hopping of excitations under imperfect spin flips.} Thus, the Stark potential can restore the DTC response even when imperfection is relatively large.

\begin{figure}[tb]
\centering
\includegraphics[width=0.48\textwidth]{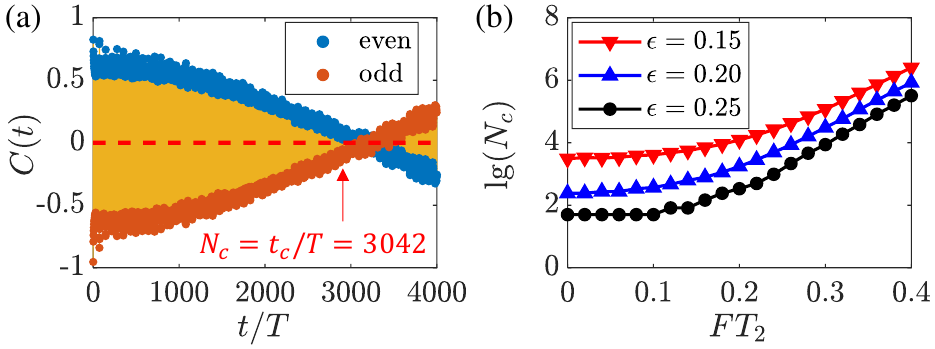}
\caption{(a) The autocorrelator $C(t)$ at stroboscopic time for $\epsilon=0.25$ and $FT_2=0.25$. The lifetime of the DTC is $N_c=t_c/T=3042$. (b) Logarithm of the lifetime $N_c$ as a function of the Stark potential strength $FT_2$ for different imperfection $\epsilon$. Other parameters are $V = 0.1$ and $L=10$.
}\label{fig4}
\end{figure}

We further take the amplitude of the subharmonic response $A_{\pi}$ to reveal the DTC order \cite{Yao2017}, with typical results shown in Fig. \ref{fig3}. We numerically calculate $A_{\pi}$ as functions of imperfection $\epsilon$ and Stark potential strength $FT_2$ with $V=0.1$ in Fig. \ref{fig3}(a). One can find that the robustness of the DTC order against imperfections is enhanced via the Stark potential and the parameter region for the DTC order is enlarged as the potential strength is increased. To be more clearly, we show $A_{\pi}$ with respect to $\epsilon$ for $FT_2=0,0.2,0.3$ in Fig. \ref{fig3}(b). We also plot $A_{\pi}$ as a function of $FT_2$ for  $V=0.06,0.09,0.12$ in Fig. \ref{fig3}(c). The results show that the Stark potential can further enhance the robustness of the DTC order with the increase of the Rydberg interaction strength $V$.

In general, the interaction of Rydberg atoms is beyond nearest neighbors. To reveal the ignorable effect of interactions beyond nearest neighbors on the DTC order, we can replace the term $\sum_{j=1}^{L-1}V n_j n_{j+1}$ with the van
der Waals form $1/2\sum_{i\neq j}^{L}V/\lvert i-j \rvert^{6}\  n_i n_{j}$ in Eqs.~\eqref{Ht}, \eqref{H1} and~\eqref{H2}, which means that the Rydberg interaction among all atoms is considered (ALL). When $\lvert i-j \rvert < 2, 3, 4$, it means nearest-neighbor (NN) interaction of the strength $V$, next-nearest-neighbor (NNN) and next-next-nearest-neighbor (NNNN) interactions are considered, respectively. {However, due to the short-range nature of the van der Waals interaction, one has minor NNN interaction $V_{\text{nnn}}=V/2^6\approx0.016V$ and NNN interaction $V_{\text{nnnn}}=V/3^6\approx0.0014V$. Thus, one can expect that the enhancement of the DTC order by the Stark potential under moderate NN interaction is not detrimental in the presence of long-range interactions.} Fig. \ref{fig3}(d) display the numerical results of $A_{\pi}$ with respect to $FT_2$ in the cases of NN, NNN, NNNN, and ALL. Compared to NN case considered previously, interactions beyond the NN term marginally affect the DTC order, and thereby one can ignore the long-range interactions.

Furthermore, we numerically find that the Stark potential can prolong the lifetime of the DTC. The amplitude of the autocorrelator $C(t)$ is observed to follow an envelope-shaped oscillation over long time scales. One can define the lifetime of the DTC $N_c$ as the stroboscopic time at which the first sign reversal in the amplitude of a Floquet cycle (either even or odd period) occurs \cite{Huang2018,Fan2020}. For instance, as shown in Fig. \ref{fig4} (a), the lifetime of the DTC is $N_c=3042$ as the first sign reversal occurs in the even period at $t/T=3042$. In Fig. \ref{fig4}(b), we show that the lifetime of the DTC increases with the increase of the Stark potential strength under different imperfections. {Notably, the double-period oscillation can persist over thousands of Floquet cycles for an array of $L=10$ Rydberg atoms, which allows the direct experimental test even for a small-size system. We also numerically find that the lifetime $N_c$ increases nearly exponentially with the system size from $L=6$ to $L=12$ (not shown here). Thus one can expect that the period-doubled DTC response enhanced by the Stark potetnial could persist in an infinite system. However, to be more accurate, a scaling analysis of the persistence of the DTC response for larger system sizes and based on the effective Hamiltonian should be numerically and analytically performed in future work, respectively.}

\begin{figure}[tb]
    \centering
	\includegraphics[width=0.46\textwidth]{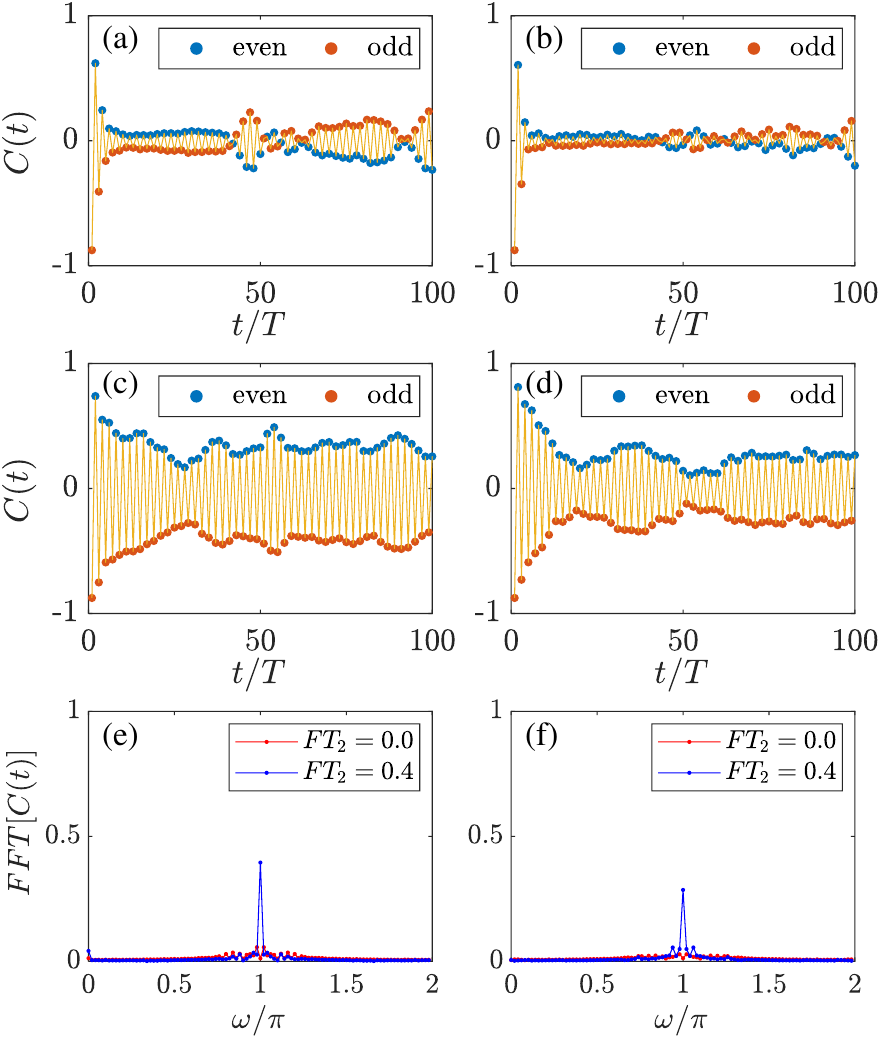}
	\caption{The autocorrelator $C(t)$ (the first and second row panels) and corresponding Fourier spectrum $FFT[C(t)]$ (the last row panels) for initial states $\ket{1111000000}$ (a,c,e) and $\ket{1111010010}$ (b,d,f), respectively. The Stark potential strength is $FT_2=0$ in (a,b) and $FT_2=0.4$ in (c,d). Other parameters are $T_1=1$, $T_2=10$, $V=0.1$, $\epsilon=0.25$ and $L=10$.
	}\label{fig5}
\end{figure}

Finally, we numerically verify that the DTC order enhanced by the Stark potential persists for different initial states. For instance, we show the numerical results of another two initial product states in Fig. \ref{fig5}. Figs. \ref{fig5}(a,b) and Figs. \ref{fig5}(c,d) present the autocorrelator dynamics in the absence and presence of the Stark potential, respectively. The corresponding Fourier spectra are shown in Figs. \ref{fig5}(e) and (f), respectively. {Apart from product states that are easily prepared in experiments, the enhancement of the DTC order is also exhibited for entangled initial states.} The independence of initial states enables one to observe the DTC order under the Stark potential without special quantum state preparation.

\section{CONCLUSION}
\label{sec:conclusion and outlook}
In summary, we have proposed a scheme to realize disorder-free DTCs in periodically driven Rydberg atom arrays with the Stark potential in the atomic detuning. We have numerically demonstrated that the Stark potential not only enhances the robustness of the DTC oder against the flip imperfections, but also extends the lifetime of the DTC, which are independent of initial states. Compared to the DTC stabilized by disorder-induced localization that requires many realizations, our scheme provides a way to explore clean DTCs in Rydberg atom arrays with much fewer resources and without special state preparation.

\begin{acknowledgments}
This work was supported by the National Natural Science Foundation of China (Grants No. 12174126, No. 12074132, and No. U20A2074), the Guangdong Basic and Applied Basic Research Foundation (Grants No. 2024B1515020018 and No. 2024A1515012516), Guangdong Provincial Quantum Science Strategic Initiative (Grant No. GDZX2303006), and the Science and Technology Program of Guangzhou (Grant No. 2024A04J3004).
\end{acknowledgments}

\normalem
\bibliography{reference}
\end{document}